\documentclass{article}
\usepackage{ijcai24}

\usepackage{times}
\usepackage{soul}
\usepackage{url}
\usepackage[hidelinks]{hyperref}
\usepackage[utf8]{inputenc}
\usepackage[small]{caption}
\usepackage{graphicx}
\usepackage{amsmath}
\usepackage{amsthm}
\usepackage{booktabs}
\usepackage{algorithm}
\usepackage{algorithmic}
\usepackage{listings}
\DeclareFixedFont{\ttb}{T1}{txtt}{bx}{n}{8.3} 
\DeclareFixedFont{\ttm}{T1}{txtt}{m}{n}{8.3}  
\usepackage{color}
\definecolor{deepblue}{rgb}{0,0,0.5}
\definecolor{deepred}{rgb}{0.6,0,0}
\definecolor{deepgreen}{rgb}{0,0.5,0}
\newcommand\pythonstyle{\lstset{
language=Python,
basicstyle=\ttm,
morekeywords={self},              
keywordstyle=\ttb\color{deepblue},
emph={MyClass,__init__},          
emphstyle=\ttb\color{deepred},    
stringstyle=\color{deepgreen},
frame=tb,                         
showstringspaces=false
}}
\lstnewenvironment{python}[1][]
{
\pythonstyle
\lstset{#1}
}
{}

\title{Validating LLM-Generated Programs with Metamorphic Prompt Testing}

\author{
Xiaoyin Wang
\and
Dakai Zhu
\affiliations
The University of Texas at San Antonio\\
\emails
\{Xiaoyin.Wang, Dakai.Zhu\}@utsa.edu
}

\begin{document}

\maketitle

\begin{abstract}

The latest paradigm shift in software development brings in the innovation and automation afforded by Large Language Models (LLMs), showcased by Generative Pre-trained Transformer (GPT), which has shown remarkable capacity to generate code autonomously, significantly reducing the manual effort required for various programming tasks. Although, the potential benefits of LLM-generated code are vast – most notably in efficiency and rapid prototyping – as LLMs become increasingly integrated into the software development lifecycle and hence the supply chain, complex and multifaceted challenges arise as the code generated from these language models carry profound questions on quality and correctness. Research is required to comprehensively explore these critical concerns surrounding LLM-generated code. 

In this paper, we propose a novel solution called metamorphic prompt testing to address these challenges. Our intuitive observation is that intrinsic consistency always exists among correct code pieces but may not exist among flawed code pieces, so we can detect flaws in the code by detecting inconsistencies. Therefore, we can vary a given prompt to multiple prompts with paraphrasing, and to ask the LLM to acquire multiple versions of generated code, so that we can validate whether the semantic relations still hold in the acquired code through cross-validation. Our evaluation on HumanEval shows that metamorphic prompt testing is able to detect 75\% of the erroneous programs generated by GPT-4, with a false positive rate of 8.6\%. 

\end{abstract}

\section{Introduction}
\label{sec:intro}
Large Language Models (LLM) have shown their capability of generating code based on natural language prompts~\cite{ni2023lever}~\cite{ouyang2023llm}~\cite{vaithilingam2022expectation}. A recent report shows that GPT-4 can create correct programs for 84\% of a set of 164 natural language prompts in HumanEval~\cite{chen2021evaluating}. 
Such capability has the potential to largely reduce manual effort in software development, but also raises severe concern on the quality and correctness of LLM-generated code. Intuitively, LLMs are trained with existing code in large code repositories such as GitHub, which may already contain bugs and flaws. Since the code generation process behind LLMs can still not be explicitly explained, it is uncertain whether they will bring in additional bugs through inconsistent synthesis of existing code. Without proper assessment of the code, developers may not confidently incorporate LLM-generated code into their project, and if they do so, the bugs hidden in the generated code may end up causing system failures and severe consequences. 

Existing techniques on model and code quality validation can hardly detect erroneous LLM-generated code pieces. Some benchmarks (e.g., HumanEval~\cite{chen2021evaluating}) and techniques (e.g., EvalPlus~\cite{liu2023your}) have been developed to evaluate the capability of LLMs on code generations, but they all rely on pre-defined canonical solutions (i.e., ground truth programs), so they can be used only for the evaluation of LLMs instead of validation of LLM-generated code in real world, because canonical solutions do not exist in the latter scenario (otherwise LLM-based code generation would be unnecessary). A commonly used solution in other AI application areas is to have a human being to double check AI products, such as have a proofreader to double check text generated by AI models. However, in the code generation scenario, source code is well known to be hard to understand, and it often takes more effort for professional developers to understand and correct other developers' code (or AI-generated code) than to write the code piece themselves~\cite{xia2017measuring}. Static source code scanners such as SpotBugs~\cite{spotbugs} and SonarQube~\cite{sonarqube} are able to automatically detect some bugs in LLM-generated code. These tools are equipped with hundreds of well-known bug patterns, so they can scan the LLM-generated code using pattern recognition to check whether it contains instances of bug patterns (e.g., calling hashcode function on array pointers, using string concatenation to form SQL queries). This solution focuses on only generic software flaws so they cannot assess task-specific semantics of LLM-generated code such as whether the generated code is consistent with the task described in the prompt. Finally, automatic testing approaches such as fuzzing~\cite{bohme2017directed} and search-based testing~\cite{mcminn2011search} can automatically generate a large number of inputs to test LLM-generated code, but due to the lack of test oracles (i.e., the ground truth output of a given input for a program, used in testing as assertions to check whether a test case passes or fails), the generated test cases can detect only run-time errors such as dead loops and crashes instead of assertion errors where the program execution completes successfully but provide an erroneous output. Our evaluation of GPT-4~\cite{adesso2022gpt4} on HumanEval (see Section~\ref{sec:expr}) shows that run-time errors are rarely triggered by LLM-generated code and all errors we observed are assertion errors. 

In this paper, to overcome the challenge of validating LLM-generated code without canonical solutions or ground-truth outputs, we propose a novel solution called \textit{metamorphic prompt testing} based on metamorphic testing~\cite{chen2018metamorphic}. Metamorphic testing is an existing software testing technique to test programs without test oracles. Its basic intuition is to take advantage of the known relations between certain inputs and outputs to check for consistency. For example, we know a relation $sin(x) == sin(\pi - x)$, so given a program $p$ that calculates the $sin$ value of an input $t$, although we do not know the ground truth oracle of $sin(t)$, we can still check whether $p(t) == p(\pi - t)$ and report an error if this is not the case. The relations been used in metamorphic testing are called $metamorphic relations$, and the major limitation of metamorphic testing (and also an important research topic in testing AI models themselves~\cite{ma2020metamorphic}~\cite{yuan2022unveiling}) is that it is typically very difficult to find a $metamorphic relation$ for an arbitrary program. Therefore, we cannot directly apply metamorphic testing to the validation of LLM-generated code. 

In \textit{metamorphic prompt testing}, our proposed new technique, instead of leveraging concrete metamorphic relations of individual LLM-generated programs, we leverage just one simple metamorphic relation: \textit{Taking prompts with the same meaning, the LLM should generate programs with the same semantics.}. Therefore, we design the basic steps of our approach as follows. First, given a prompt, we use NLP-based paraphrasing~\cite{witteveen2019paraphrasing}~\cite{shahmohammadi2021paraphrase} to create multiple paraphrases of the prompt, referred to as \textit{paraphrase prompts}. Second, we feed the original prompt to the LLM-based code generator and generate the original LLM-generated program. This program is the one we would like to validate and we refer to it as the \textit{target program}. Third, we feed all paraphrase prompts to the LLM-based code generator and generate a set of LLM-generated programs which we referred to as the \textit{paraphrase programs}. Fourth, we use the automatic test generation technique~\cite{bohme2017directed} to create random inputs to feed into the target program as well as the paraphrase programs and check whether their outputs are identical. It should be noted that, in this last step, to address the noises brought in by the paraphrasing and code generation process, we do not require all outputs to be identical, but developed an algorithm to resolve conflicts. 

We evaluated metamorphic prompt testing on the well known benchmark HumanEval~\cite{chen2018metamorphic}. Our experiment results show that, among 164 code generation prompts in HumanEval, our approach is able to averagely detect 75\% (potentially 83\% with a more conservative configuration) out of all 24 erroneous programs generated by GPT-4, with a false positive rate of 8.6\%. In other words, a developer need to look at only 30 LLM-generated programs to find 18 of them as truly erroneous (with 12 false positives), while missing only 6 erroneous programs. If adopting our approach as a post-processor of GPT-4 by rejecting the generated programs identified as erroneous, it is able to boost the accuracy from 85.4\% to 89.0\%. 

To sum up, our paper makes the following contributions.

\begin{itemize}
    \item We developed a novel approach to validate LLM-generated programs without the requirement of any canonical solutions or ground truth output. 
    \item We developed a novel algorithm to report errors based on conflicting output of multiple LLM-generated programs from paraphrased prompts. 
    \item We report an experiment to evaluate our approach on the well know benchmark HumanEval. 
\end{itemize}

The remaining of this paper is organized as follows. In Section~\ref{sec:background}, we will introduce some background knowledge of AI-based code generation and metamorphic testing. In Section~\ref{sec:approach}, we will describe the detailed steps of our approach. In Section~\ref{sec:expr}, we will present the research questions to be answered in our experiment, our experiment setup, and results. After pointing to some future works in Section~\ref{sec:future}, we conclude in Section~\ref{sec:conclusion}.

\section{Background}
\label{sec:background}

\subsection{AI-based Code Generation}
Artificial Intelligence (AI)-based code generation is a very recent advancement in the field of AI to support software development. It has the potential fundamentally altering how programmers, from novices to experts, approach coding tasks. 

AI-based code generation starts from early efforts on API method suggestion~\cite{huang2018api} and code auto-completion~\cite{cruz2021automated} based on traditional machine learning models~\cite{liu2018effective} and later deep neural network~\cite{cruz2021automated}. More recently, with the notion that programs are also to some extent natural~\cite{hindle2016naturalness}, researchers start to train large code models~\cite{nguyen2018deep}, or incorporating large amount of code when training general-purpose LLMs~\cite{adesso2022gpt4}. At the heart of AI-based code generation is the ability of these systems to interpret a wide range of natural language inputs, from simple commands to complex problem descriptions. The LLMs then processes this input, applying its extensive knowledge base and understanding of various programming languages and paradigms, to produce syntactically correct and logically sound code. This not only accelerates the coding process but also helps in reducing human error~\cite{zhang2021autotrainer}, leading to more efficient and reliable software development. 

One of the key benefits of AI in code generation is its accessibility~\cite{jonsson2022cracking}. It empowers individuals with limited coding experience to develop software solutions, and can be easily used for education purposes~\cite{becker2023programming}. For experienced developers, it may acts as an assistant to automating repetitive coding and offering suggestions for optimization and debugging, like in co-pilot~\cite{nguyen2022empirical}. The major limitation of AI-based code generation is that they rely heavily on the quality of input provided and their training data, which can sometimes lead to erroneous programs as output. Since developers cannot tell which output program is erroneous, and errors in programs may lead to severe consequences, this uncertainty becomes the biggest obstacle preventing developers from confidently use AI-generated code. Our paper is moving toward a potential solution of this problem.

\subsection{Metamorphic Testing}

Software testing is a critical phase in the software development life cycle, aiming at ensuring the reliability and correctness of software applications. Traditional testing methods rely heavily on test oracles, which are mechanisms for determining whether a software application behaves as expected for a given set of inputs. However, in many real-world scenarios, especially in complex systems like machine learning models, scientific computations, or data analytics applications, knowing the exact expected output can be challenging or even impossible. 

Metamorphic Testing~\cite{chen2018metamorphic} is a technique that overcomes the oracle problem by leveraging metamorphic relations — properties or rules about how the output should change in response to specific changes in the input. Instead of looking for exact output values, it focuses on how the output changes when inputs are modified in a controlled manner.

For instance, consider a navigation app that calculates the shortest path between two points. It might be difficult to know the exact correct path in every scenario, but one can reasonably expect that if the destination is moved closer to the source, the path length should decrease. This expectation forms a metamorphic relation. Metamorphic testing typically consists of the following phases. 

\begin{itemize}
    \item Identification of Metamorphic Relations: The first step involves identifying properties or rules that define how the software's output should change in response to changes in the input. These relations are central to metamorphic testing and are derived from the underlying logic of the application.
    \item Generation of Test Cases: Once metamorphic relations are established, test cases are generated by altering the original input in ways that the metamorphic relations dictate.
    \item Verification of Output Changes: The software is then executed with both the original and altered inputs, and the outputs are compared. The key here is not to verify the correctness of the output itself but to check if the changes in the output adhere to the metamorphic relations.
\end{itemize}

Due to the ability of metamorphic testing to handle unknown output, it has been adopted in testing AI models, which also have non-deterministic output. For example, Ma et al.~\cite{ma2020metamorphic} developed a metamorphic=testing-based approach to test NLP models and detect potential unfair biases. However, as mentioned in Section~\ref{sec:intro} and above, metamorphic testing has the main limitation that it needs the identification of metamorphic relations, which often require human effort in a case-by-case solution. Therefore it cannot be directly applied to LLM-generated programs which are arbitrary. In our approach, we solve this issue by lifting the metamorphic testing to the prompt level.

\section{Approach}
\label{sec:approach}
\begin{figure*}[h]
    \centering
    \includegraphics[width=\textwidth]{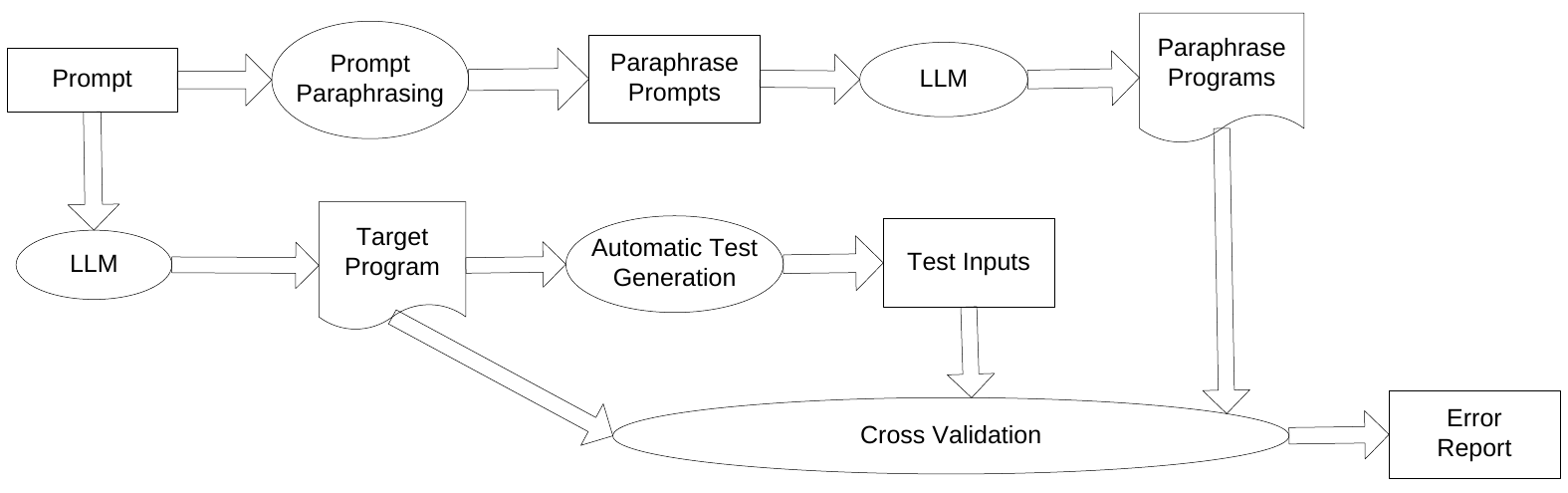}
    \caption{AR App bugs}
    \label{fig:overview}
\end{figure*}

The overview of our approach is illustrated in Figure~\ref{fig:overview}. From the figure, we can see that our approach takes just a prompt as the input and does not need any additional information. On the one hand, the prompt is fed into a prompt paraphrasing component to generate a set of paraphrase prompts. On the other hand, the prompt is fed into the LLM to create a target program. The paraphrase prompts are also fed into the LLM to create paraphrase programs. After that, we apply automatic test generation tool to the target program to create a set of test inputs (without test oracles). The target program, the paraphrase programs, and the test inputs are then combined in the cross validation module which report whether the target program is erroneous or not. Except for LLM which is considered a black box in our approach, the other three major components are the prompt paraphrasing, automatic test generation, and the cross validation. We will introduce each of them in detail in the following subsections. 

\begin{python}[language=python, caption={Paraphrases of Prompt No. 9}, label={list:paraphrase}]
from typing import List, Tuple
def rolling_max(numbers: List[int]) -> List[int]:
    """
From a given list of integers, generate a list 
of rolling maximum element found until given 
moment in the sequence.
>>> rolling_max([1, 2, 3, 2, 3, 4, 2])
    [1, 2, 3, 3, 3, 4, 4]
    """
------------------------------------
from typing import List, Tuple
def rolling_max(numbers: List[int]) -> List[int]:
    """
Create a list that details the highest integer 
present up to each point within a provided 
sequence of integers.
>>> rolling_max([1, 2, 3, 2, 3, 4, 2])
    [1, 2, 3, 3, 3, 4, 4]
    """
------------------------------------
from typing import List, Tuple
def rolling_max(numbers: List[int]) -> List[int]:
    """
Create a list that reflects the highest number 
encountered so far from a sequence of 
integers as you progress through it.
>>> rolling_max([1, 2, 3, 2, 3, 4, 2])
    [1, 2, 3, 3, 3, 4, 4]
    """
\end{python}

\subsection{Generation of Paraphrase Programs}
We illustrate how we generate paraphrase programs using the example in List~\ref{list:paraphrase}. The first section (above the first separation line) of the example shows an original prompt in HumanEval. Each prompt provides a semi-structured natural language description of a function, and consists of the following parts. 

\begin{itemize}
    \item The import statements which declare the python library types to be used, as shown in the first line of each section of List~\ref{list:paraphrase}.
    \item The signature of the function to be implemented, including the types of inputs of outputs, as shown in the second line of each section of List~\ref{list:paraphrase}.
    \item The natural language description of the semantics of the function to be implemented, as shown in Lines 4-6 of List~\ref{list:paraphrase}.
    \item Examples showing some sample inputs and their corresponding outputs, as shown in Lines 7 and 8 of each section of List~\ref{list:paraphrase}.
\end{itemize}
    
Since we need to run all the paraphrase programs with the same set of test inputs, we need to make sure that they all have the same signature. Meanwhile, the examples need to be left identical because any changes on them will lead to a semantic change. Therefore, when creating paraphrases of prompts, we consider only the description part of a prompt, while skip all the other three parts of the prompt and leave them as untouched. To create paraphrase of the description part, we use the same LLM model, and feed a prompt by adding ``Can you paraphrase the following paragraph?'' before the description part of the prompt. 

Once paraphrases of the description part are provided by the LLM, we replace the original description part with the paraphrases to create the paraphrase prompts as shown in the second and third section of List~\ref{list:paraphrase}. Finally, we feed the original prompt and the paraphrase prompts into the LLM to acquire the target program and the paraphrase programs. We add ``Can you generate python code for the following function?'' before the prompts when feeding them to the LLM to make sure the LLM is creating implementations rather than tests or comments. 

\subsection{Automatic Test Generation}
In the earlier step, we acquired the target program and a set of paraphrase programs, and we need to determine which program we should create test cases for. Since our main goal is to validate the target program, we want to make sure the automatically generated test inputs exercise most of its paths and achieve high code coverage on it. Therefore, we consider only the target program when generating test inputs. 

In our approach, conceptually we can adopt any automatic test generation tool, so our approach can be enhanced together with more advanced test generation tools. In our implementation, we use the python-AFL tool~\cite{pythonafl} (python-oriented version of AFL, or American Fuzzy Lop~\cite{fioraldi2023dissecting}) to create test inputs because it is very robust and uses the state-of-the-practice coverage-guided fuzzing technique. In traditional fuzz testing, random data is input into a program to uncover errors or vulnerabilities. Coverage-guided fuzzing enhances fuzz testing by incorporating feedback from the program during testing. In particular, it monitors the program to see which parts of the code are being exercised (the code coverage) by the test inputs. This information is then used to generate new test input (through mutation of test input from earlier rounds) that targets unexplored areas of the program, leading to more thorough testing coverage. Python-AFL is very fast, but to make sure our test set is comparable with the original test sets in HumanEval, we limit the number of our generated test inputs to 20 because the former typically has less than 20 test inputs. 

\subsection{The Cross Validation Algorithm}
After we acquire the set of test inputs, we will run the target program and the paraphrase programs with them and cross validate the results. A most conservative strategy would be reporting an error whenever a different output is seen (i.e., when any paraphrase program's output on any input is different from that of the target program). While this can be considered as a valid configuration, especially for critical usage scenarios, we believe such strict consistency may not be necessary given that noises may be brought in during the paraphrase generation process (e.g, an erroneous paraphrase may have a different meaning from the original prompt). 

To achieve a more balanced cross validation, we developed Algorithm~\ref{alg:algorithm}. The basic idea is to report an error only if the majority of the paraphrase programs have different semantics from the target program (i.e., generating a different output for any test input). The algorithm takes three inputs: the target program ($Target$), the set of paraphrase programs ($Paraprogs$), and the set of test inputs ($Tests$). The output of the algorithm is a boolean flag. In the algorithm, we execute each paraphrase programs with each input (shown in Line 6) and compare the results with output of the target program. If there is any difference, we will add the corresponding paraphrase program into the $DiffSet$. Finally, we check whether the size of $DiffSet$ is more than half of the number of paraphrase programs.

\begin{algorithm}[tb]
    \caption{Cross Validation Algorithm}
    \label{alg:algorithm}
    \textbf{Input}: Target, Paraprogs, Tests\\
    \textbf{Output}: ErrorFlag
    \begin{algorithmic}[1] 
        \STATE Let $ErrorFlag=False$
        \STATE Let $DiffSet=\emptyset$
        \FOR{$Test$ $\in$ $Tests$}
        \STATE Let $Output$ = $run(Target, Test)$
        \FOR{$Para$ $\in$ $Paraprogs$}
        \IF {$run(Para, Test) != Output$}
        \STATE Add $Para$ to $DiffSet$
        \ENDIF
        \ENDFOR
        \ENDFOR
        \STATE \textbf{return} $len(DiffSet) > len(Paraprogs) / 2$
    \end{algorithmic}
\end{algorithm}

\section{Experiments}
\label{sec:expr}
In this section, we will introduce the evaluation results of our approach on the HumanEval benchmark. 
\subsection{Research Questions}
In our evaluation, we try to answer the following research questions. 
\begin{itemize}
    \item \textbf{RQ1:} How effective is our approach on detecting erroneous LLM-generated code?
    \item \textbf{RQ2:} How our approach compares with its variants on detecting erroneous LLM-generated code (Ablation Study)?
    \item \textbf{RQ3:} What are the characters of erroneous LLM-generated code that our approach detected correctly and incorrectly? 
\end{itemize}

\subsection{Evaluation Metrics}
To measure the effectiveness of our approach, we use the following widely adopted metrics: accuracy, recall, and false positive rate. Since all these metrics require definitions of true / false positives and true / false negatives, we define them in our application scenario as follows. 

\begin{itemize}
    \item True Positives (TP): Since our goal is to detect erroneous LLM-generated programs, a truly erroneous LLM-generated program reported by our approach as erroneous is considered as a true positive.
    \item False Positives (FP): A correct LLM-generated program reported by our approach as erroneous is considered as a true positive. 
    \item True Negatives (TN): A correct LLM-generated program reported by our approach as also correct is considered as a true negative. 
    \item False Negatives (FN): A truly erroneous program reported by our approach as correct is considered as a false negative. 
\end{itemize}

Based on the definitions above, the metrics we use are defined as follows. Among these metrics, the accuracy mainly measures the overall performance of our approach. The recall mainly measures the how likely our approach is able to detect an erroneous LLM-generated program, and the false positive rate / precision mainly measures how likely our approach is giving a false alarm that wastes developers' time and effort. 

\begin{equation}
    Accuracy = \frac{TP + TN}{TP + FP + TN + FN}
\end{equation}
\begin{equation}
    Precision = \frac{FP}{FP + TP}
\end{equation}
\begin{equation}
    Recall = \frac{TP}{TP + FN}
\end{equation}
\begin{equation}    
    False Positive Rate = \frac{FP}{FP + TN}
\end{equation}

\subsection{Evaluation Setup}
We evaluate our approach on the widely adopted HumanEval dataset~\cite{chen2021evaluating}. The dataset consists of 164 pairs of human written prompts and corresponding canonical solutions as ground truth. It also provides a set of test cases for each prompt so that a given LLM-generated program can be evaluated against the canonical solution. It should be noted that we use neither canonical solutions nor test cases in our approach. We use them only for evaluation of our approach. For the LLM, we used the \texttt{gpt-4-1106-preview} model provided by OpenAI because GPT-4 is the state-of-the-art model for code generation~\cite{muennighoff2023octopack}, and the model we use is its most recent version. 

\subsection{Experiment Results}
To answer \textbf{RQ1}, for each prompt, we check the LLM-generated target program against the canonical solution with test cases provided by HumanEval to acquire the ground truth label of the target program. Then we compare the reports of our approach with the ground truth labels to calculate the metrics. The results are presented in Table~\ref{table:357}. 

\begin{table}
    \centering
    \begin{tabular}{crrrr}
        \hline
        \#prompts  & Accuracy & Recall & Precision & FP Rate \\
        \hline
        3          &     88.4\%    &  58.3\%    &  60.9\%      &      6.5\%   \\
        5          &     89.0\% &    75.0\% &    60.0\%   &  8.6\%       \\
        7          &     86.6\%  &   75.0\%  &   52.9\%  &   11.4\%  \\
        \hline
    \end{tabular}
    \caption{Effectiveness of Our Approach with Different Number of Prompts}
    \label{table:357}
\end{table}

In the table, Column 1 shows the number of prompts we used. Since we can create different number of paraphrase prompts, we would like study how that may affect our evaluation results. We consider odd numbers only here so that our cross validation always works with a clear majority. Columns 2-5 present the accuracy, recall, precision, and FP rate, respectively. 

From the table, we have the following observations. First of all, the best configuration of our approach (\#prompts = 5) is able to averagely detect 75\% (potentially 83\% with a more conservative configuration) out of all 24 erroneous programs generated by GPT-4, with a false positive rate of 8.6\%. In other words, a developer need to look at only 30 LLM-generated programs to find 18 of them as truly erroneous (with 12 false positives), while missing only 6 erroneous programs. Second, the configuration with 3 prompts has much lower recall and lower false positive rate as well. When the number of prompts is small, due to the lower variety of prompts, the approach may have less power to create different paraphrase program variants, making it more difficult to detect erroneous programs. Third, the configuration with 7 prompts achieves the same recall as the configuration with 5 prompts, but with more false positives, so its precision and FP rate are lower. The reason behind may be that more prompts may lead to more noises in the paraphrase generation process (i.e. paraphrases generated by LLM may also not be true paraphrases). 

To answer \textbf{RQ2}, we perform an ablation study to find out whether the techniques in our approach are useful. We particularly want to evaluate the effectiveness of our prompt paraphrasing technique, so we compare our approach with an alternative that does not perform paraphrasing, but just feed the same prompt to the LLM for multiple times. Since we use five prompts as the default configuration of our approach, we also feed the same prompt five times in this variant. Another technique we want to evaluate is our cross-validation algorithm. Here in the variant we consider the more straightforward conservative cross validation, where we report an error whenever we see a different output from any paraphrase program on any input. 

\begin{table}
    \centering
    \begin{tabular}{crrrr}
        \hline
        Variant  & Accuracy & Recall & Precision & FP Rate \\
        \hline
        Our Approach   &     89.0\% &    75.0\% &    60.0\%   &  8.6\%       \\
        w/o Paraphra.   &     87.2\%    &  58.3\%    &  56.5\%      &      7.1\%   \\

        conservative           &    77.4\%  &   83.3\%  &   36.7\%  &   22.1\%  \\
        cross valid.       &             &            &         &               \\
        \hline
    \end{tabular}
    \caption{Effectiveness of Our Approach with Different Number of Prompts}
    \label{table:ablation}
\end{table}

The evaluation results are presented in Table~\ref{table:ablation}. In the table, Column 1 shows the name of the variant, and Columns 2-5 present the accuracy, recall, precision, and FP rate, respectively. From the table, we have the following observations. First, compared with our approach, the variant without paraphrasing achieves lower false positive rate, but its recall is also much lower. Since repetitively feeding in the same prompt may not trigger more variance in the code generation process, the newly generated code is more likely to be identical or very similar to the target program. Therefore, this variant is not as effective as our approach on detecting erroneous LLM-generated program. Meanwhile, it also brings in fewer false positives. Second, the variant using conservative cross validation achieves higher recall, but its precision is much lower. Since the conservative cross validation reports strictly more errors than our default approach, this observation is expected. The precision of 36.7\% indicate that a developer may need to review almost 2 false positives when detecting one true erroneous programs. However, the additional effort may worth it in critical scenarios because this variant does find two 8.3\% more erroneous programs. Furthermore, the false positive rate of 22.1\%, although higher than all other variants and configurations of our approach, still indicate that 78\% of the programs have been filtered out so a developer may not need to review them. 

\subsection{Qualitative Analysis}
To answer \textbf{RQ3}, we manually analyzed the failing cases (false positives and false negatives) of our approach. We will summarize their characteristics and showcase some examples. Our manual analysis shows that the major reason for false negatives of our approach lies in the high similarity among paraphrases, as shown in List~\ref{list:fnexample}. Note that we show only two paraphrase prompts here for briefness. From the List, we can see that the two paraphrase prompts generated (in parts 2 and 3) are very similar to the original prompt (in part 1). The lack of variety in paraphrase prompts caused the LLM to generate very similar paraphrase programs and thus all of them share the same wrong semantics. 

\begin{python}[language=python, caption={Paraphrases of Prompt No. 83}, label={list:fnexample}]
def starts_one_ends(n):
"""
For a given positive integer n, calculate the 
total quantity of positive integers with n digits 
that either begin or conclude with the digit 1.
"""
------------------------------------
def starts_one_ends(n):
"""
For a given positive integer n, calculate the 
total amount of n-digit positive integers that 
either begin or conclude with the digit 1.
"""
------------------------------------
def starts_one_ends(n):
"""
For a positive integer n, calculate the total 
number of n-digit positive integers that either 
begin or conclude with the digit 1.
"""
\end{python}

Our manual analysis of false positives show that the major reason lies in the paraphrase clustering effect, as shown in List~\ref{list:fpexample}. From List~\ref{list:fpexample}, we can see that the paraphrase prompts (in part 2 and part 3) arse different from the original prompt (in part 1). However, the similarity among the paraphrase prompts themselves are high (note that here we show only two paraphrase prompts and the remaining paraphrase prompts are also very similar to the two shown here). Since all the paraphrase prompts look very similar to each other, once the LLM generate erroneous code for one of them, it will also generate erroneous code for all of them, and thus these erroneous code easily holds a majority among prompts and cause our approach to report a false positive. 

From these observations, we can see that the similarity among generated paraphrase prompts plays an important role in the effectiveness of our approach. Therefore, it is possible that we can measure the similarity to indicate whether our approach is applicable or not in certain cases. It may also be possible to try to enlarge the variance of generated paraphrase prompts to further enhance our approach. 

\begin{python}[language=python, caption={Paraphrases of Prompt No. 26}, label={list:fpexample}]
from typing import List
def remove_duplicates(numbers: List[int]):
""" 
From a list of integers, remove all elements 
that occur more than once. Keep order of elements 
left the same as in the input.
    """
------------------------------------
from typing import List
def remove_duplicates(numbers: List[int]):
    """
Eliminate any repeating integers from a sequence 
while maintaining the original order of the 
remaining numbers.
    """
------------------------------------
from typing import List
def remove_duplicates(numbers: List[int]):
    """
Eliminate any duplicate integers from the sequence, 
ensuring that the remaining elements retain their 
original order as presented in the input.
    """
\end{python}

\subsection{Threats to Validity}
The threats to the construction validity measure whether our experiment setting is the same as real-world usage of our approach. Our approach takes only a prompt as its input, and our experiment setup does the same way. The major threats to the internal validity of our evaluation is the potential bugs and errors in our implementation of our approach, variants, and scripts to collect and analyze data. To reduce such threats, we carefully checked all of our code and scripts during experiment and run experiments for multiple times to ensure consistency. The major threats to the external validity of our evaluation is that we evaluated our approach on only GPT-4 and HumanEval, which consists of 164 prompts for python code generation. It is possible that our evaluation results may not apply to other LLMs, datasets, and programming languages. To reduce this threat, we use the state-of-the-art LLM and the most widely adopted benchmark to perform our evaluation. To further reduce this threat, we plan to perform more evaluations on more LLMs and more benchmarks in different programming languages in the future. 

\section{Future work}
\label{sec:future}
The research results presented in this paper shows that using metamorphic prompt testing to validate LLM-generated programs is a promising direction. We plan to further extend our research toward this direction in the following aspects. First of all, we plan to extend the scope of our experiments to run more studies on more benchmarks from different programming languages and considering more LLMs beyond GPT-4. Second, we plan to investigate how paraphrase prompt similarity may affect the effectiveness of our approach and develop novel techniques to take advantage of that. We may be able to predict whether our approach is applicable based on whether the generated paraphrase prompts are too similar. We may also be able to develop techniques to enlarge the difference between paraphrase prompts to reduce false positives and false negatives of our approach. Third, we plan to explore metamorphic relations other than prompt paraphrasing. We can further explore metamorphic relations such as negation to enlarge the set of prompts we can use to validate LLM-generated code. Fourth, since our approach is able to detect LLM-generated code errors, it may be used to provide feedback to the LLM or code generation models or be used to fine tune LLM-based code generation models. 

\section{Conclusion}
\label{sec:conclusion}
Large language models or LLMs have show strong ability to generate code with high accuracy. However, due to the complexity of source code and severe consequences may be caused by code errors, this code-generation capability can hardly be adopted by developers in real world because they do not have the mechanism to validate the LLM-generated code with relatively low effort. 
In this paper, we propose a novel solution called metamorphic prompt testing to address this challenge. Our solution is based on metamorphic testing and varies the prompts to create paraphrase prompts, and then cross validate LLM-generated programs of multiple prompts. Our evaluation on HumanEval shows that metamorphic prompt testing is able to detect 75\% of the erroneous programs generated by GPT-4, with a false positive rate of 8.6\%. In our experiment, we try to answer three research questions and we can draw the following conclusions. 

\begin{itemize}
    \item Metamorphic prompt testing is shown by an evaluation on HumanEval as an effective approach to validate LLM-generated code. It can achieve high recall with low false negative rate. Therefore, developer may need to examine very small subset of LLM-generated code and find most of the errors. 
    \item As shown by our ablation study, our proposed techniques prompt paraphrasing and cross validation algorithms are both useful, enhancing the effectiveness of our approach. 
    \item Similarity among paraphrase prompts is an important factor leading to false positives and false negatives of our approach. More research is required to further take advantage of this observation. 
\end{itemize}



\bibliographystyle{named}
\bibliography{ijcai24}

\begin{thebibliography}{}

\bibitem[\protect\citeauthoryear{Adesso}{2022}]{adesso2022gpt4}
Gerardo Adesso.
\newblock Gpt4: The ultimate brain.
\newblock {\em Authorea Preprints}, 2022.

\bibitem[\protect\citeauthoryear{Becker \bgroup \em et al.\egroup
  }{2023}]{becker2023programming}
Brett~A Becker, Paul Denny, James Finnie-Ansley, Andrew Luxton-Reilly, James
  Prather, and Eddie~Antonio Santos.
\newblock Programming is hard-or at least it used to be: Educational
  opportunities and challenges of ai code generation.
\newblock In {\em Proceedings of the 54th ACM Technical Symposium on Computer
  Science Education V. 1}, pages 500--506, 2023.

\bibitem[\protect\citeauthoryear{B{\"o}hme \bgroup \em et al.\egroup
  }{2017}]{bohme2017directed}
Marcel B{\"o}hme, Van-Thuan Pham, Manh-Dung Nguyen, and Abhik Roychoudhury.
\newblock Directed greybox fuzzing.
\newblock In {\em Proceedings of the 2017 ACM SIGSAC conference on computer and
  communications security}, pages 2329--2344, 2017.

\bibitem[\protect\citeauthoryear{Chen \bgroup \em et al.\egroup
  }{2018}]{chen2018metamorphic}
Tsong~Yueh Chen, Fei-Ching Kuo, Huai Liu, Pak-Lok Poon, Dave Towey, TH~Tse, and
  Zhi~Quan Zhou.
\newblock Metamorphic testing: A review of challenges and opportunities.
\newblock {\em ACM Computing Surveys (CSUR)}, 51(1):1--27, 2018.

\bibitem[\protect\citeauthoryear{Chen \bgroup \em et al.\egroup
  }{2021}]{chen2021evaluating}
Mark Chen, Jerry Tworek, Heewoo Jun, Qiming Yuan, Henrique Ponde de~Oliveira
  Pinto, Jared Kaplan, Harri Edwards, Yuri Burda, Nicholas Joseph, Greg
  Brockman, et~al.
\newblock Evaluating large language models trained on code.
\newblock {\em arXiv preprint arXiv:2107.03374}, 2021.

\bibitem[\protect\citeauthoryear{Cruz-Benito \bgroup \em et al.\egroup
  }{2021}]{cruz2021automated}
Juan Cruz-Benito, Sanjay Vishwakarma, Francisco Martin-Fernandez, and Ismael
  Faro.
\newblock Automated source code generation and auto-completion using deep
  learning: Comparing and discussing current language model-related approaches.
\newblock {\em AI}, 2(1):1--16, 2021.

\bibitem[\protect\citeauthoryear{Fioraldi \bgroup \em et al.\egroup
  }{2023}]{fioraldi2023dissecting}
Andrea Fioraldi, Alessandro Mantovani, Dominik Maier, and Davide Balzarotti.
\newblock Dissecting american fuzzy lop: a fuzzbench evaluation.
\newblock {\em ACM Transactions on Software Engineering and Methodology},
  32(2):1--26, 2023.

\bibitem[\protect\citeauthoryear{Hindle \bgroup \em et al.\egroup
  }{2016}]{hindle2016naturalness}
Abram Hindle, Earl~T Barr, Mark Gabel, Zhendong Su, and Premkumar Devanbu.
\newblock On the naturalness of software.
\newblock {\em Communications of the ACM}, 59(5):122--131, 2016.

\bibitem[\protect\citeauthoryear{Huang \bgroup \em et al.\egroup
  }{2018}]{huang2018api}
Qiao Huang, Xin Xia, Zhenchang Xing, David Lo, and Xinyu Wang.
\newblock Api method recommendation without worrying about the task-api
  knowledge gap.
\newblock In {\em Proceedings of the 33rd ACM/IEEE International Conference on
  Automated Software Engineering}, pages 293--304, 2018.

\bibitem[\protect\citeauthoryear{Jonsson and
  Tholander}{2022}]{jonsson2022cracking}
Martin Jonsson and Jakob Tholander.
\newblock Cracking the code: Co-coding with ai in creative programming
  education.
\newblock In {\em Proceedings of the 14th Conference on Creativity and
  Cognition}, pages 5--14, 2022.

\bibitem[\protect\citeauthoryear{Liu \bgroup \em et al.\egroup
  }{2018}]{liu2018effective}
Xiaoyu Liu, LiGuo Huang, and Vincent Ng.
\newblock Effective api recommendation without historical software
  repositories.
\newblock In {\em Proceedings of the 33rd ACM/IEEE International Conference on
  Automated Software Engineering}, pages 282--292, 2018.

\bibitem[\protect\citeauthoryear{Liu \bgroup \em et al.\egroup
  }{2023}]{liu2023your}
Jiawei Liu, Chunqiu~Steven Xia, Yuyao Wang, and Lingming Zhang.
\newblock Is your code generated by chatgpt really correct? rigorous evaluation
  of large language models for code generation.
\newblock {\em arXiv preprint arXiv:2305.01210}, 2023.

\bibitem[\protect\citeauthoryear{Ma \bgroup \em et al.\egroup
  }{2020}]{ma2020metamorphic}
Pingchuan Ma, Shuai Wang, and Jin Liu.
\newblock Metamorphic testing and certified mitigation of fairness violations
  in nlp models.
\newblock In {\em IJCAI}, pages 458--465, 2020.

\bibitem[\protect\citeauthoryear{McMinn}{2011}]{mcminn2011search}
Phil McMinn.
\newblock Search-based software testing: Past, present and future.
\newblock In {\em 2011 IEEE Fourth International Conference on Software
  Testing, Verification and Validation Workshops}, pages 153--163. IEEE, 2011.

\bibitem[\protect\citeauthoryear{Muennighoff \bgroup \em et al.\egroup
  }{2023}]{muennighoff2023octopack}
Niklas Muennighoff, Qian Liu, Armel Zebaze, Qinkai Zheng, Binyuan Hui,
  Terry~Yue Zhuo, Swayam Singh, Xiangru Tang, Leandro Von~Werra, and Shayne
  Longpre.
\newblock Octopack: Instruction tuning code large language models.
\newblock {\em arXiv preprint arXiv:2308.07124}, 2023.

\bibitem[\protect\citeauthoryear{Nguyen and Nadi}{2022}]{nguyen2022empirical}
Nhan Nguyen and Sarah Nadi.
\newblock An empirical evaluation of github copilot's code suggestions.
\newblock In {\em Proceedings of the 19th International Conference on Mining
  Software Repositories}, pages 1--5, 2022.

\bibitem[\protect\citeauthoryear{Nguyen \bgroup \em et al.\egroup
  }{2018}]{nguyen2018deep}
Anh~Tuan Nguyen, Trong~Duc Nguyen, Hung~Dang Phan, and Tien~N Nguyen.
\newblock A deep neural network language model with contexts for source code.
\newblock In {\em 2018 IEEE 25th International Conference on Software Analysis,
  Evolution and Reengineering (SANER)}, pages 323--334. IEEE, 2018.

\bibitem[\protect\citeauthoryear{Ni \bgroup \em et al.\egroup
  }{2023}]{ni2023lever}
Ansong Ni, Srini Iyer, Dragomir Radev, Veselin Stoyanov, Wen-tau Yih, Sida
  Wang, and Xi~Victoria Lin.
\newblock Lever: Learning to verify language-to-code generation with execution.
\newblock In {\em International Conference on Machine Learning}, pages
  26106--26128. PMLR, 2023.

\bibitem[\protect\citeauthoryear{Ouyang \bgroup \em et al.\egroup
  }{2023}]{ouyang2023llm}
Shuyin Ouyang, Jie~M Zhang, Mark Harman, and Meng Wang.
\newblock Llm is like a box of chocolates: the non-determinism of chatgpt in
  code generation.
\newblock {\em arXiv preprint arXiv:2308.02828}, 2023.

\bibitem[\protect\citeauthoryear{{Python-AFL}}{}]{pythonafl}
{Python-AFL}.
\newblock {Python-AFL}.
\newblock \url{https://github.com/jwilk/python-afl}.

\bibitem[\protect\citeauthoryear{Shahmohammadi \bgroup \em et al.\egroup
  }{2021}]{shahmohammadi2021paraphrase}
Hassan Shahmohammadi, MirHossein Dezfoulian, and Muharram Mansoorizadeh.
\newblock Paraphrase detection using lstm networks and handcrafted features.
\newblock {\em Multimedia Tools and Applications}, 80:6479--6492, 2021.

\bibitem[\protect\citeauthoryear{{SonarQube}}{}]{sonarqube}
{SonarQube}.
\newblock {SonarQube}.
\newblock \url{https://www.sonarsource.com/products/sonarqube/}.

\bibitem[\protect\citeauthoryear{{SpotBugs}}{}]{spotbugs}
{SpotBugs}.
\newblock {SpotBugs}.
\newblock \url{https://spotbugs.github.io/}.

\bibitem[\protect\citeauthoryear{Vaithilingam \bgroup \em et al.\egroup
  }{2022}]{vaithilingam2022expectation}
Priyan Vaithilingam, Tianyi Zhang, and Elena~L Glassman.
\newblock Expectation vs. experience: Evaluating the usability of code
  generation tools powered by large language models.
\newblock In {\em Chi conference on human factors in computing systems extended
  abstracts}, pages 1--7, 2022.

\bibitem[\protect\citeauthoryear{Witteveen and
  Andrews}{2019}]{witteveen2019paraphrasing}
Sam Witteveen and Martin Andrews.
\newblock Paraphrasing with large language models.
\newblock {\em arXiv preprint arXiv:1911.09661}, 2019.

\bibitem[\protect\citeauthoryear{Xia \bgroup \em et al.\egroup
  }{2017}]{xia2017measuring}
Xin Xia, Lingfeng Bao, David Lo, Zhenchang Xing, Ahmed~E Hassan, and Shanping
  Li.
\newblock Measuring program comprehension: A large-scale field study with
  professionals.
\newblock {\em IEEE Transactions on Software Engineering}, 44(10):951--976,
  2017.

\bibitem[\protect\citeauthoryear{Yuan \bgroup \em et al.\egroup
  }{2022}]{yuan2022unveiling}
Yuanyuan Yuan, Qi~Pang, and Shuai Wang.
\newblock Unveiling hidden dnn defects with decision-based metamorphic testing.
\newblock In {\em Proceedings of the 37th IEEE/ACM International Conference on
  Automated Software Engineering}, pages 1--13, 2022.

\bibitem[\protect\citeauthoryear{Zhang \bgroup \em et al.\egroup
  }{2021}]{zhang2021autotrainer}
Xiaoyu Zhang, Juan Zhai, Shiqing Ma, and Chao Shen.
\newblock Autotrainer: An automatic dnn training problem detection and repair
  system.
\newblock In {\em 2021 IEEE/ACM 43rd International Conference on Software
  Engineering (ICSE)}, pages 359--371. IEEE, 2021.

\end{thebibliography}

\end{document}